\documentclass[a4paper,12pt]{scrartcl}

\RequirePackage{xr}
\RequirePackage[OT1]{fontenc}
\RequirePackage{amsthm,amsmath,amssymb,dsfont,bm}
\RequirePackage{algorithm}
\RequirePackage{algpseudocode}
\RequirePackage{graphicx} 
\RequirePackage[section]{placeins}
\RequirePackage{color}
\RequirePackage{chngcntr}
\RequirePackage{enumerate}
\usepackage[shortlabels]{enumitem}
\RequirePackage{longtable}
\usepackage{booktabs}
\usepackage{array}
\usepackage{multirow}
\usepackage{mathtools}
\usepackage[english]{babel} 
\usepackage{xcolor}
\RequirePackage[OT1]{fontenc} \RequirePackage{amsthm}
\RequirePackage{amsfonts} \RequirePackage{amssymb}
\RequirePackage{bm}

\RequirePackage{graphicx}
\RequirePackage{verbatim}
\usepackage{url}

\numberwithin{equation}{section}
\RequirePackage{a4wide} 
\theoremstyle{plain}

\newtheorem{theorem}{Theorem}[section]

\newtheorem{remark}{Remark}[section]

\def\@bysame#1{\vrule height 1.5pt depth -1pt width 3em \hskip
0.5em\relax}

\newcommand{\R}{\mathbb{R}}

\newcommand{\eat}[1]{}

\newcommand{\matD}{\bm  D}
\newcommand{\matid}{\bm  I}
\newcommand{\vecnull}{\boldmath 0}

\newcommand{\bfSigma}{\boldsymbol{\Sigma}}

\renewcommand{\phi}{{\scriptsize \varphi}}%{{\scriptsize \varphi}}

\DeclareMathOperator*{\argmin}{arg\,min}

\DeclareMathOperator{\Var}{\mathbb Var}
\DeclareMathOperator{\Cov}{\mathbb Cov}

\DeclareMathOperator{\E}{\mathbb{E}}

\newcommand{\Z}{ \mathbb{Z} }

\newcommand{\trunc}[1]{ {\lfloor #1 \rfloor} }
\newcommand{\wh}[1]{ \widehat{ #1 } }
\newcommand{\wt}[1]{ \widetilde{ #1 } }

\newcommand{\calD}{\mathcal{D}}

\newcommand{\calF}{\mathcal{F}}

\newcommand{\calM}{\mathcal{M}}
\newcommand{\calS}{\mathcal{S}}

\newcommand{\calU}{\mathcal{U}}

\newcommand{\calV}{\mathcal{V}}

\newcommand{\matA}{{\bm A}}
\newcommand{\matB}{{\bm B}}
\newcommand{\matC}{{\bm C}}

\newcommand{\matH}{{\bm H}}

\newcommand{\matT}{{\bm T}}

\newcommand{\matW}{{\bm W}}

\newcommand{\veca}{{\bm a}}

\newcommand{\vecB}{{\bm B}}
\newcommand{\vecc}{{\bm c}}

\newcommand{\vece}{{\bm e}}

\newcommand{\vecv}{{\bm v}}

\newcommand{\vecw}{{\bm w}}

\newcommand{\vecY}{{\bm Y}}

\newcommand{\bfLambda}{\bm\Lambda}

\usepackage{subfigure}
\usepackage{mathtools} % ab hier selber eingebunden
\usepackage{dsfont}
\usepackage{color}
\usepackage{amsthm}

\usepackage{float}                 % für erweiterten Symbolvorrat

\begin{document}

\begin{center}
	\begin{minipage}{.8\textwidth}
		\centering 
		\LARGE Shrinkage for Covariance Estimation: Asymptotics, Confidence Intervals, Bounds and Applications in Sensor Monitoring and Finance \\[0.5cm]

		\normalsize
		\textsc{Ansgar Steland}\\[0.1cm]
		Institute of Statistics,\\
		RWTH Aachen University,\\
		Aachen, Germany\\
		Email: \verb+steland@stochastik.rwth-aachen.de+
		
	\end{minipage}
\end{center}

\begin{abstract}
	When shrinking a covariance matrix towards (a multiple) of the identity matrix, the trace of the covariance matrix arises naturally as the optimal scaling factor for the identity target. The trace also appears in other context, for example when measuring the size of a matrix or the amount of uncertainty.
	Of particular interest is the case when the dimension of the covariance matrix is large. Then the problem arises that the sample covariance matrix is singular if the dimension is larger than the sample size. Another issue is that usually the estimation has to based on correlated time series data. We study the estimation of the trace functional allowing for a high-dimensional time series model, where the dimension is allowed to grow with the sample size - without any constraint. Based on a recent result, we investigate a confidence interval for the trace, which also allows us to propose lower and upper bounds for the shrinkage covariance estimator as well as bounds for the variance of projections. In addition, we provide a novel result dealing with shrinkage towards a diagonal target. 
	
	We investigate the accuracy of the confidence interval by a simulation study, which indicates good performance, and analyze three stock market data sets to illustrate the proposed bounds, where the dimension (number of stocks) ranges between $32$ and $475$. Especially, we apply the results to portfolio optimization and determine bounds for the risk associated to the variance-minimizing portfolio. 
\end{abstract}


%\textit{MSC 2010 subject classifications: } Primary 62F12, 62H30, Secondary . 

\section{Introduction}
\label{sec: Introduction}

In diverse fields such as finance, natural science or medicine the analysis of high-dimensional time series data is of increasing importance. In the next section, we consider data from financial markets and sensor arrays, for instance consisting of photocells (solar cells), as motivating examples for high-dimensional data. Here the number of time series, the dimension $d$, can be much larger than the sample size $n$. Then standard assumptions such as $ d $ fixed and $ n \to \infty $, the classical low-dimensional setting, or $ d/n \to y \in (0,1) $, as in random matrix theory, \cite{RMT}, are not justifiable. Even when $ d < n $, so that - theoretically - the covariance matrix may have nice properties such as invertability, it is recommended to regularize the sample covariance matrix when $d$ is large. A commonly used method is shrinkage as studied in depth by \cite{LW2003}, \cite{LW2004} and for weakly dependent time series in \cite{Sanc2008}, among others. Here the trace functional of the sample covariance matrix arises as a basic ingredient for shrinkage. The trace also arises in other settings, e.g. as the trace norm $ \| \matA \|_{tr} = \text{tr}( \matA ) $  to measure the size of a nonnegative definite matrix $ \matA $, or when measuring the total information. For the latter application, recall that the variance $ \sigma^2 $ of a zero mean random variable $X$ with finite second moment is a natural measure of the uncertainty of $ X $ and a canonical measure of its precision is $ 1/\sigma^2 $. For $d$ random variables it is natural to consider the total variance defined as the sum of their variances. These applications motivate us to study estimators for the trace and, especially, their statistical evaluation in terms of variance estimators and confidence intervals, lower and upper bounds; the problem of regularized covariance estimation solved by (linear) shrinkage represents our key application.

We study variance estimators and a related easy-to-use confidence interval for the trace of a covariance matrix, when estimating the latter by time series data. By \cite{StelandSachs2017b}, the estimator is asymptotically normal and the variance estimator turns out to be consistent under a high-dimensional framework, which even allows that the dimension $ d = d_n $  grows in an arbitrary way, as $ n \to \infty $. Indeed, the results of \cite{StelandSachs2017b}, which are based on \cite{StelandSachs2017a}, and those of the present paper do not require any condition on the dimension and the sample size such as $ d_n/n \to \zeta \in (0,1) $, contrary to results using random matrix theory. 
 
These results allow us to construct an easy-to-calculate confidence interval, which in turn allows us to make inference and to quantify the uncertainty associated to the proposed estimator in a statistically sound way. The results also suggest novel lower and upper data-based bounds for the shrinkage covariance estimator, and these bounds in turn yield lower and upper data-based bounds for the  variance of a projection of the $d$-dimensional observed vector onto a projection vector.  We evaluate the confidence interval by its real coverage probability and examine its accuracy by a simulation study for high dimensions. Here we consider settings where the dimension is up to $50$ times larger than the length of the time series.

Going beyond the identity target for shrinkage covariance estimation, this paper also contributes new asymptotic results when shrinking towards a diagonal matrix. Concretely, we consider case of a diagonal target corresponding to uncorrelated coordinates. Then the shrinkage covariance estimator strengthens the diagonal of the sample covariance matrix.  Our results deal with a strong approximation by a Gaussian random diagonal matrix. Again, the result holds true without any constraint on the dimension.  

The paper is organized as follows. Motivating applications to finance and sensor monitoring, which also lead to the assumed high-dimensional model framework, are discussed in Section~\ref{ExamplesAndAssumptions}. In Section~\ref{ref: sec1}, we review the trace functional and discuss its role for shrinkage estimation. Section~\ref{ref: est} provides the details about the proposed variance estimator and the asymptotic large sample approximations.  Especially, Section~\ref{sec: asymptotics_trace}  reviews the estimator proposed and studied in  \cite{StelandSachs2017a} based on the work of \cite{StelandSachs2017b} and discusses the proposed confidence interval. A new result about the diagonal shrinkage target is provided in Section~\ref{ref: new}. Simulations and the application to financial stock market data are presented in Section~\ref{sec: sims}. Our application especially covers portfolio optimization as one of the most important problems related to investment. As well known, the strategy of the portfolio selection process heavily determines the risk associated to the portfolio return. Here we follow the classical approach to measure risk by the variance and consider the variance-minimizing portfolio calculated from a shrinkage covariance estimator. Our results provide in a natural way lower and upper bounds for the portfolio risk. We illustrate their application by analyzing three data sets of stock market log returns. 

\section{Motivating example and assumptions}
\label{ExamplesAndAssumptions}

Let us consider the following motivating examples.

\subsection{High-dimensional sensor monitoring} 

Suppose a source signal is monitored by a large number $d$ of sensors. The source $ \{ \epsilon_k : k \in \Z \} $ is assumed to be given by independent zero mean noise with possibly heterogenous (finite) variances,
\[
	\epsilon_k \sim (0, \sigma_k^2), \qquad \text{{\em independent}}, 
\]
for constants $ \sigma_k^2 \ge 0 $, $ k \in \Z $. Here we write $ X \sim (a,b) $ for a random variable $X$ and constants $a \in \R $ and $ b \ge 0 $, if $ X $ follows an arbitrary distribution with  mean $a$ and existing variance $b$.  

The above model implies that the information present in the source signal is coded in the variances $ \sigma_k^2 $. The source is in a homogeneous or stable state, if it emits a signal with constant variance. Let us consider the testing problem given by the  null hypothesis of homogeneity,
\begin{center}
	$ H_0 : $ The source signal is i.i.d., $ \sigma_k^2 = \textcolor{black}{s_0^2} $, for all $ k \in \Z $.
\end{center}
If the source emits a signal with a non-constant variance, we may say that it is in an unstable state. This can be formulated, for instance, by the alternative hypothesis
\begin{center}
	$ H_1 : $ $ \{ \epsilon_k \} $  are independent with variances $ \{ \sigma_k^2 \}  $ satisfying $ \sum_{i=1}^n (\sigma_i^2 - \overline{\sigma^2}_n )^2 > 0 $ for $n \ge 2$,
\end{center}
where $ \overline{\sigma^2}_n = \frac{1}{n} \sum_{i=1}^n \sigma^2_i $. Note that $ H_1 $ represents the complement of $ H_0 $ within the assumed class of distributions for $ \{ \epsilon_k : k\in \Z \} $ with independent, zero mean coordinates under moment conditions specified later.
Depending on the specific application, certain patterns may be of interest, of course, and demand for specialized procedures, but this issue is beyond the scope of this paper. Let us, however, briefly discuss a less specific situation frequently studied, namely when a change-point occurs where the uncertainty of the signal changes: If at a certain time instant, say $q $, the variance changes to another value, $q$ is called change-point and we are led to a change-point alternative,
\begin{center}
	$H_{1,cp}^{(q)} :   \sigma_q^2 = \textcolor{black}{s_1^2} \not= s_0^2 = \sigma_{\ell} $, for $ \ell < q $.
\end{center}

Let us now assume that the source is monitored by $d$ sensors which deliver to a central data center a flow of possibly correlated discrete measurements in the form of a time series. Denote by  $ Y_t^{(\nu)} $ the real-valued measurement of the $ \nu$th sensor received at time $ t $, $ t = 1, 2, \dots $. We want to allow for sensor arrays which are possibly spread over a large area and therefore receive the source signal at different time points. Further, we have in mind sensors which aggregate the input over a certain time frame, such as capacitors, Geiger counters to detect and measure radiation or the photocells of a camera sensor. Therefore, let us make the following assumption about the data-processing mechanism of the sensors:
\begin{itemize}
	\item The sensor $ \nu $ receives the source signal with a delay $ \delta_\nu \ge 0 $, such that $ \epsilon_{t-\delta_\nu} $ (instead of $ \epsilon_t $) influences $ Y_t^{(\nu)} $: \[ \epsilon_{t-\delta_\nu} \to Y_t^{(\nu)} \]
	\item Previous observations $  \epsilon_{t-j} $, $ j > \delta_\nu $, affect the sensor, but they are damped by weights $ c_j^{(\nu)} $
	\[ \epsilon_{t-j} \stackrel{c_j^{(\nu)}}{\to} Y_t^{(\nu)} \]
\end{itemize}
This model can also be justified by assuming that the source signal may be disturbed and reflected, e.g. at buildings etc., such that at a certain location we cannot receive $ \epsilon_k $ but only a mixture of that current signal value and past observations.

These realistic assumptions call for the well known concept of a linear filter providing the output, such that a natural model taking into account the above facts is to assume that the time series $ Y_t^{(\nu)} $ available for statistical analyses follow a linear process,
\begin{equation}
\label{ModelTS}
Y_t^{(\nu)} = \sum_{j=0}^\infty c_j^{(\nu)} \epsilon_{t-j}, \qquad t = 1, 2, \dots,  \nu = 1, \dots, d.
\end{equation}
Lastly, it is worth mentioning that sensor devices frequently do not output raw data but apply signal processing algorithms, e.g. low and/or high pass filters, which also result in outputs following (\ref{ModelTS}), even if one can observe $ \epsilon_t $ at time $t$. For example, image sensors use built-in signal processing to reduce noise, enhance contrast or, for automotive applications, detect lanes, see \cite{HsiaoEtAl2009}. For body sensor networks in health monitoring 3D acceleration signals need to be filtered to maximize the signal-to-noise ratio. \cite{WangEtAl2011} develop and study a 3D sensor with a built-in Butterworth low-pass filter with waveform delay correction. 

\subsection{Financial time series}

Linear time series are also a common approach to model econometric and financial data such as log return series of assets defined as 
\[
  r_t = \log P_t / P_{t-1}
\]
where $ P_t $ denotes the time $t$ price of a share. Although the serial correlations of the daily log returns of a single asset are usually quite small or negligible, the cross correlations between different assets are relevant and are used to reduce investment risk by proper diversification. Hence, models such as (\ref{ModelTS}) for $d$ series of daily log returns are justifiable. Extensions to factor models are preferable; they are  subject of current research and will be published elsewhere.

Instead of analyzing marginal moments, analyzing conditional variances of log returns by means of GARCH models and their extensions has become quite popular, and we shall briefly review such models to clarify and discuss the differences to the class of models studied in the present paper. 

Recall that $ \{ e_t : t \in \Z  \} $ is called a GARCH($p,q$)-process, see \cite{Bol1986}, \cite{FZ} and \cite{Ste},  if it is a martingale difference sequence with respect to the natural filtration $ \calF_t = \sigma( e_s : s \le t ) $, and if there exist constants $ \omega, \alpha_1, \dots, \alpha_q, \beta_1, \dots, \beta_p $ such that the conditional variance $ \wt{\sigma}_t^2 $ satisfies the equations
\[
\wt{\sigma}_t^2 = \Var( e_t | \calF_{t-1} ) = \omega + \sum_{i=1}^q \alpha_i e_{t-i}^2 + \sum_{j=1}^p \beta_j \wt{\sigma}_{t-j}^2, \qquad t \in \Z.
\]
For conditions on the parameters ensuring existence of a solution we refer to \cite{FZ}, see also \cite[Theorem~3.7.6]{Ste}. 
Such GARCH processes show volatility clusters which is one of the reasons of their success in financial modelling.
Putting $ \nu_t = e_t^2 - \wt{\sigma}_t^2 $ and substituting the $ \wt{\sigma}_{t-j}^2 $ by $ e_{t-j}^2 - \nu_{t-j} $ it follows that the squares $ \epsilon_t^2 $ can be written as
\[
\epsilon_t^2 = \omega + \sum_{i=1}^r (\alpha_i + \beta_i) e_{t-i}^2 + \nu_t - \sum_{j=1}^p \beta_j \nu_{t-j}, \qquad t \in \Z,
\]
where $ r = \max(p,q) $, $ \alpha_i = 0 $ if $ i >q $ and $ \beta_j = 0 $ if $ j > p $, and $ \nu_t = e_t^2 - \sigma_t^2 $ are the innovations. This equation shows that the squares of a GARCH($p,q$) process follow an ARMA($r,p$)-process with respect to the innovations $ \nu_t $. The GARCH approach can therefore be interpreted as an approach which analyzes the conditional mean of the second moments by an ARMA model, i.e. as a (linear) function of the information set $ \calF_t $. Various extensions, such as the exponential GARCH etc., have been studied, which consider different models for the conditional variance in terms of the information set.

Consider now a  zero mean $d$-dimensional time series  $ \vece_t $ and let $ \bfSigma_t = \E( \vece_t \vece_t' ) $ and $ \matH_t = \E( \vece_t \vece_t'  | \calF_t) $, where now $ \calF_t = \sigma( \vece_s : s \le t ) $. Multivariate extensions of the GARCH approach model the matrix of the conditional second moments, $ \matH_t $, as a function of the information set $ \calF_t $. For example, the so-called vec representation considers the model
\[
  \text{vech}( \matH_t ) = \matW + \sum_{j=1}^q \matA_j \text{vech}( \vece_{t-j} \vece_{t-j}' ) + \sum_{j=1}^p \matB_j \matH_{t-j}
\]
for coefficient matrices $ \matW, \matA_1, \dots, \matA_q, \matB_1, \dots, \matB_p $, see \cite{EK} and \cite{FZ}. Here the vector-half operator $ \text{vech}(\matA) $ stacks the $ d(d+1)/2 $ elements of the lower triangular part of a matrix $ \matA$. Whereas this model is designed to analyze the conditional variances and covariances of the coordinates $ e_{t}^{(\nu)} $ of $ \vece_t $, which determine the marginal second moments, modelling the centered squares  $ ( e_t^{(\nu)} )^2 - \E(e_t^{(\nu)} )^2  $ by (\ref{ModelTS}), such that 
\begin{equation}
\label{ModelSquares}
	( e_t^{(\nu)} )^2 - \E(e_t^{(\nu)} )^2 = \sum_{j=0}^\infty c_j^{(\nu)} \epsilon_{t-j}, \qquad 1 \le \nu \le d, t \in \Z,
\end{equation}
models the dependence structure of the squares  $ ( e_t^{(\nu)} )^2 $ and implies the model
\[
  \left( \Cov( [e_t^{(\nu)} ]^2, [e_t^{(\mu)} ]^2 ) \right)_{1 \le \nu \le d \atop 1 \le j} = \matC \bfLambda \matC'
\]
for their covariance matrix, where  
\[
  \matC = \left( c_{nj}^{(\nu)}  \right)_{1 \le \nu \le d \atop 1 \le j}, \qquad \bfLambda = \text{diag}( \sigma_0^2, \sigma_1^2, \cdots )
\]
with $ \sigma_j^2 = \E( \eta_j^2 ) $ for $ j \ge 1 $. We may write
	\[
\Var( \vecY_{ni} ) 	= \sum_{j=0}^\infty \sigma_{i-j}^2 \vecc_{nj} \vecc_{nj}'
= \sigma_{i}^2 \vecc_{n0} \vecc_{n0}' + \sigma_{i-1}^2 \vecc_{n1} \vecc_{n1}'  + \cdots 
\]
with $ \vecc_{nj} = ( c_{nj}^{(1)}, \dots, c_{nj}^{(d_n)} )' $. Therefore, in this model hypotheses dealing with inhomogeneity of $ \Var( \vecY_{ni} ) $, $ i = 1, \dots, n $,  may be a consequence of a change in the variances $ \sigma_j^2 $, or result from a change of the coefficients summarized in the vectors $ \vecc_{nj} $.

\subsection{Assumptions}

The theoretical results used below and motivated above assume model (\ref{ModelTS}) and require the following conditions.

\vskip 0.2cm
\noindent
\textbf{Assumption 1:} The  innovations $ \epsilon_t $ have finite absolute moments of the order $ 8 $.

\vskip 0.1cm
\noindent
\textbf{Assumption 2:} The coefficients satisfy the decay condition
\begin{equation}
\label{DecaySharp}
\sup_{n \ge 1} \max_{1 \le \nu  \le d_n} | c_{nj}^{(\nu)} | \le C j^{-(1+\delta')},
\end{equation}
for some $ \delta' > 0 $. This condition is weak enough to allow for ARMA($p,q$) models, \[ \phi(L) X_t = \theta(L) \epsilon_t, \] where $ \phi(L) $ is a lag polynomial of order $p$ and $ \theta(L) $ a lag polyonmial of order $q$.  It is worth mentioning that also seasonal ARMA models with $s$ seasons are covered, where the observations $ X_{j + s t} $, $ t = 1, 2, \dots $, of season $j$ follows an ARMA($p,q$) process,  \[ \Phi( B^s ) X_t = \Theta( B^s ) \epsilon_t, \] for lag polynomials $ \Phi $ and $ \Theta $, see e.g. \cite{BrockwellDavis} for details.

\section{The trace functional and shrinkage} 
\label{ref: sec1}
 
 Let $ \bfSigma $ be the covariance matrix of a zero mean random vector $ \vecY = (Y^{(1)}, \dots, Y^{(d)} )' $ of dimension $d$.  Recall that the trace of $ \bfSigma $ is defined as the sum of the diagonal,
 \[
  \text{tr}( \bfSigma ) = \sum_{\nu=1}^d \Var( Y^{(\nu)} ).
 \]
The related average,
\[
	\text{tr}^*( \bfSigma ) = d^{-1} \text{tr}( \bfSigma )
\]
is called {\em scaled trace}. Observe that it assigns the value $ \text{tr}^*( \matid ) = 1 $ to the identity matrix, $ \matid $, whereas $ \text{tr}( \matid ) = d \to \infty $, if the dimension tends to infinity.
The trace resp. scaled trace arises naturally in the form of a scaling factor when shrinking the true covariance matrix $ \bfSigma $ towards the identity matrix under the Frobenius norm, see \cite{LW2003} and \cite{LW2004}, which represents the simplistic model of uncorrelated, homogenous coordinates. 

Denote by $ \calM $ the set of $d \times d $ matrices and denote by $ \calS $ the subset of covariance matrices of dimension $d \times d $. Equip $ \calM $ with the inner product
\[
  (\matA, \matB) = \text{tr}(\matA'\matB), \qquad \matA, \matB \in \calM,
\]
which induces the Frobenius matrix norm $ \| \matA \|_F = \sqrt{ (\matA, \matA) } $, $ \matA \in \calM $. Then $ \calM $ becomes a separable Hilbert space of dimension $d^2$.  The orthogonal projector, $ \Pi : \calM \to \calU $, onto the one-dimensional linear subspace
\[
  \calU = \text{span} \{ \matB \} = \{ \lambda \matB : \lambda \in \R \}
\]
associated to a single matrix $ \matB \not= \vecnull $ is given by
\[
  \Pi( \matA; \matB ) = \frac{ (\matA, \matB) \matB }{ (\matB, \matB) }.
\]
Clearly, $ \Pi( \matA; \matB ) $ is the optimal element from $ \calU $ which minimizes the distance \[ d( \matA, \calU ) = \inf \{ d(\matA, \matB) : \matB \in \calU \} \]
between $ \matA $ and the subspace  $ \calU $: $ \Pi(\matA; \matB ) $ is the element from $ \calU $ to approximate $ \matA $ best. It follows that for $ \matB = \matid $ the optimal approximation of $ \bfSigma $ by a multiple of the identity matrix is given by
\[
	\matT := \Pi( \bfSigma; \matid ) = \frac{ (\bfSigma, \matid) \matid }{ (\matid, \matid) } = d^{-1} \text{tr}( \bfSigma ) \matid.
\]
This is the optimal target for shrinking: If one wants to 'mix in' a regular matrix, then one should use $ \matT = tr^*( \bfSigma ) $. The shrunken covariance matrix with respect to a {\em shrinkage weight} $W \in [0, 1] $, also called {\em mixing parameter} or {\em shrinkage intensity}, is now defined by the convex combination
\[
  \bfSigma^s = (1-W) \bfSigma + W \Pi( \bfSigma; \matid ) = (1-W) \bfSigma + W \text{tr}^*( \bfSigma ) \matid.
\]
To summarize, the optimal shrinkage target is given by $ \text{tr}^*( \bfSigma ) \matid $ where the optimal scaling factor $ \text{tr}^*( \bfSigma )  $ is called {\em shrinkage scale}.

Provided we have a (consistent) estimator $ \widehat{\text{tr}^*( \bfSigma )} $ of $ \text{tr}^*( \bfSigma ) $, we can estimate  the shrunken covariance matrix by the shrinkage covariance estimator
\begin{equation}
\label{ShrinkEst}
  \widehat{\bfSigma}_n^s = (1-W) \widehat{\bfSigma}_n + W \widehat{\text{tr}^*( \bfSigma )} \matid,
\end{equation}
where 
\[
	\widehat{\bfSigma}_n = \frac{1}{n} \sum_{i=1}^n \vecY_{i} \vecY_{i}' 
\]
is the usual sample covariance matrix. Whatever the shrinkage weight, the shrinkage covariance estimator has 
several appealing properties: Whereas $ \widehat{\bfSigma}_n $ is singular if $ d \ge n $, the shrinkage estimator $ \widehat{\bfSigma}_n^s $ is always positive definite and thus invertible. From a practical and computational point of view, it has the benefit that it is fast to compute.  We shall, however, see that its statistical evaluation by a variance estimator is computationally more demanding. As shown in \cite{LW2003} and \cite{LW2004}, the shrinkage estimator has further optimality properties, whose discussion goes beyond the scope of this brief review. For extensions of those studies to weakly dependent time series see \cite{Sanc2008}. There it is also shown how one can select the shrinkage weight in an optimal way, if there is no other guidance.

\section{Nonparametric estimation of the scaled trace}
\label{ref: est}

In practice, one has to estimate the shrinkage target $ \matT = \text{tr}^*( \bfSigma ) \matid $, i.e. we have to estimate the scaled trace of $ \bfSigma $. Let us assume that for each coordinate $ Y^{(\nu)} $ of the vector $ \vecY $  a time series of length $n$,
\[
  Y_i^{(\nu)}, \qquad i = 1, \dots, n,
\]
is available for estimation. Put $ \vecY_i = ( Y_i^{(1)}, \dots, Y_i^{(d)} )' $, $ i = 1, \dots, n $. The canonical nonparametric estimator for $ \sigma_\nu^2 = \Var( Y^{(\nu)} ) $ is the sample moment
\[
  \widehat{\sigma}_\nu^2 = \frac{1}{n} \sum_{i=1}^n ( Y_i^{(\nu)} )^2, \qquad \nu = 1, \dots d,
\]
which suggests the plug-in estimator
\[
  \widehat{\text{tr}(\bfSigma)} = \sum_{\nu=1}^d \widehat{\sigma}_\nu^2.
\]
Obviously, we have the relationship 
\[
\widehat{\text{tr}(\bfSigma) } = \text{tr}( \widehat{\bfSigma}_n ).
\]
The scaled trace is now estimated by
\[
	\widehat{\text{tr}^*(\bfSigma) } = \text{tr}^*( \widehat{\bfSigma}_n ) = \frac{1}{d} \sum_{\nu=1}^d \widehat{\sigma}_\nu^2.
\]

\subsection{Variance estimation: uncorrelated case}

If the time series $ \{  Y_i^{(\nu)} :  i = 1, \dots, n \} $, $ \nu = 1, \dots, d $, are independent and if $ d $ is fixed, then the statistical evaluation of the uncertainty associated with the estimator $ \widehat{\text{tr}^*(\bfSigma) }  $, on which we shall focus in the sequel, is greatly simplified, since then
\begin{equation}
\label{VarTraceEstIID}
  \Var( \widehat{\text{tr}^*(\bfSigma) } ) = \frac{1}{d^2} \sum_{\nu=1}^d \Var( \widehat{\sigma}_\nu^2 ),
\end{equation}
and we may estimate this expression by estimating the $d$ variances $ \Var( \widehat{\sigma}_\nu^2 ) $, $ \nu = 1, \dots, d$ . 
Let us first stick to that case. Suppose that all time series are strictly stationary with finite absolute moments of the order $ 4+\delta $ for some $ \delta > 0 $. Then a straightforward calculation shows that
\[
  \Var( \widehat{\sigma}_\nu^2 ) = \frac{1}{n} \left[  n \gamma_\nu(0) + 2 \sum_{h=1}^{n-1} (n-h) \gamma_\nu(h)
    \right],
\]
where
\[
  \gamma_\nu(h) = \Cov( (Y_1^{(\nu)})^2, (Y_{1+|h|}^{(\nu)})^2 ), \qquad h \in \Z,
\]
is the lag $ h $ autocovariance of the squared time series.
The canonical sample autocovariance estimates
\[
	\widehat{\gamma}_\nu(h) = \frac{1}{n} \sum_{i=1}^{n-|h|}[ (Y_i^{(\nu)})^2 - \wh{\mu}_\nu][ (Y_{i+|h|}^{(\nu)})^2 - \wh{\mu}_\nu ]
\]
where $ \wh{\mu}_\nu = \frac{1}{n} \sum_{i=1}^n (Y_i^{(\nu)})^2 $,  lead to the Bartlett-type long-run variance estimator
\[
  \widehat{\Var}( \widehat{\text{tr}(\bfSigma) } )  = \widehat{\gamma}_\nu(0) + 2 \sum_{|h| \le m} w_{mh} \widehat{\gamma}_\nu( h ).
\]
Here $ w_{nh} $ are weights satisfying the usual conditions,
\begin{itemize}
\item[(i)]  $ |w_{nh} | \le W $ for some constant $W$ and 
\item[(ii)] $ w_{mh} \to 1 $, as $ m \to \infty $, for all $h$. 
\end{itemize}
Starting with \cite{NW1987} and \cite{Andrews1991} conditions under which such estimators are consistent are well known. Essentially, one has to require that the lag truncation sequence satisfies $ m \to \infty $ and $ m^2/n \to 0 $. For a result on almost sure convergence under weak conditions we refer to \cite{BerkesKokoszka2005}.
Since the estimator (\ref{VarTraceEstIID}) sums up a finite number of such estimators, the consistency easily carries over.

\subsection{Variance estimation: correlated case}

In the sequel, we want to relax two crucial conditions made above: We will now consider correlated time series and allow that the dimension $d$ depends on $n$ and may grow with $n$: $ d_n \to \infty $. Our exposition follows \cite{StelandSachs2017b}.
But if the $d$ time series are correlated, then, in general, formula (\ref{VarTraceEstIID}) no longer applies. Instead we have
\[
  \sigma_{tr}^2 = \Var( \widehat{\text{tr}^*(\bfSigma) } ) = \frac{1}{d_n^2} \sum_{\nu=1}^{d_n} \sum_{\mu=1}^{d_n} \Cov( \widehat{\sigma}_\nu^2, \widehat{\sigma}_{\mu}^2 ).  
\]
In what follows, we assume that $ \inf_{n \ge 1} \sigma_{tr}^2 > 0 $. A direct calculation reveals the long-run variance structure
\[
\beta_n^2(\nu, \mu) = \Cov( \widehat{\sigma}_\nu^2, \widehat{\sigma}_{\mu}^2 ) = \gamma_n^{(\nu,\mu)}(0) + 2 \sum_{\tau=1}^{n-1} \frac{n-\tau}{n} \gamma_n^{(\nu,\mu)}( \tau ),
\]
where
\[
  \gamma_n^{(\nu,\mu)}( \tau ) = \Cov( (Y_{1}^{(\nu)})^2, (Y_{1+|\tau|}^{(\mu)})^2 )
\]
are the lag $\tau $ cross-covariances of the squares. They can be estimated by
\[
	\widehat{\gamma}_n^{(\nu,\mu)}( \tau ) = \frac{1}{n} \sum_{i=1}^{n-|\tau|} [ (Y_{i}^{(\nu)} )^2 - \widehat{\mu}_n(\nu) ] [ (Y_{i+|\tau|}^{(\mu)} )^2 - \widehat{\mu}_n(\mu) ].
\]
Now we can estimate the covariances $ \beta_n^2(\nu, \mu)$  by the long-run variance estimators
\begin{equation}
\label{BetaEst}
  \widehat{\beta}_n^2(\nu, \mu) = \widehat{\gamma}_n^{(\nu,\mu)}(0) +  2 \sum_{\tau=1}^{m} w_{m\tau} \widehat{\gamma}_n^{(\nu,\mu)}( \tau ),
\end{equation}
for $ 1 \le \nu, \mu \le d_n $, where $ m = m_n $ is a sequence of lag truncation constants. Eventually, we are led to the estimator
\[
  \widehat{\sigma}_{tr}^2 = \frac{1}{d_n^2} \sum_{\nu, \mu=1}^{d_n^2}  \widehat{\beta}_n^2(\nu, \mu).
\]

\subsection{Asymptotics for the trace estimator}
\label{sec: asymptotics_trace}

In  \cite{StelandSachs2017b}  the asymptotics of the estimator $ \text{tr}^*( \widehat{\bfSigma}_n ) $ has been studied in depth. Let us briefly review these results. 

\begin{theorem} Suppose (\ref{ModelTS}) and Assumptions 1 and 2 hold. Then the scaled trace norm is asymptotically normal in the sense that, provided the probability space is rich enough to carry an additional uniformly distributed random variable, there exists a Gaussian random variable
\[
  Z \sim N( 0, \sigma_{tr}^2 ) 
\]
such that 
\begin{equation}
\label{GaussApprox}
  | \sqrt{n}[ \text{tr}^*( \widehat{\bfSigma}_n ) - \text{tr}^*( \bfSigma_n ) ] - Z | \to 0,
\end{equation}
as $n \to \infty $, a.s.. Further, the estimator $ \widehat{\sigma}_{tr}^2 $ for $ \sigma_{tr}^2$ is $ L_1 $-consistent, i.e.
\[
  \E | \widehat{\sigma}_{tr}^2 - \sigma_{tr}^2 | \to 0,
\]
as $n \to \infty $, if the lag truncation sequences satisfies
\[
  m_n \to \infty, \quad m_n^2 / n \to 0, 
\]
as $ n \to \infty $.
\end{theorem}

Based on the above result one may propose the confidence interval
\[
  \left[ \text{tr}^*( \widehat{\bfSigma}_n ) - z_{1-\alpha/2} \frac{\widehat{\sigma}_{tr}}{\sqrt{n}}, \text{tr}^*( \widehat{\bfSigma}_n ) + z_{1-\alpha/2} \frac{\widehat{\sigma}_{tr}}{\sqrt{n}} \right]
\]
where $ z_p $ denotes the $ p $-quantile of the standard normal distribution, i.e. $ \Phi(z_{p} ) = p  $ for $ p \in (0,1) $, where $ \Phi $ denotes the distribution function of the standard normal distribution.

For the shrinkage estimator $ \widehat{\bfSigma}_n^s $, see (\ref{ShrinkEst}), the above result allows us to calculate lower and upper bounds: A lower bound is given by
\begin{align*}
 \widehat{\bfSigma}_{n,L}^s & = (1-W) \widehat{\bfSigma}_n + W \left( \text{tr}^*( \widehat{\bfSigma}_n ) -   z_{1-\alpha/2} \frac{\widehat{\sigma}_{tr}}{\sqrt{n}} \right) \matid \\
   & = \widehat{\bfSigma}_n^s - W z_{1-\alpha/2} \frac{\widehat{\sigma}_{tr}}{\sqrt{n}} \matid,
\end{align*}
and an upper bound by
\begin{align*}
\widehat{\bfSigma}_{n,U}^s &= (1-W) \widehat{\bfSigma}_n + W \left( \text{tr}^*( \widehat{\bfSigma}_n ) + z_{1-\alpha/2} \frac{\widehat{\sigma}_{tr}}{\sqrt{n}} \right) \matid \\
   & = \widehat{\bfSigma}_n^s + W z_{1-\alpha/2} \frac{\widehat{\sigma}_{tr}}{\sqrt{n}} \matid.
\end{align*}
Observe that these bounds differ only on the diagonal. From a statistical point of view, they provide the {\em justifiable} minimal and maximal amount of strengthening of the diagonal of the sample covariance matrix.

Suppose now that we estimate the variance \[ \sigma_n^2(\vecw_n) = \Var( \vecw_n' \vecY_n ) = \vecw_n' \bfSigma_n \vecw_n \] of the projection $ \vecw_n' \vecY_n $ onto a projection vector $ \vecw_n $ with uniformly bounded $ \ell_1 $-norm using the shrinkage covariance estimator
\[
  \widehat{\Var}( \vecw_n' \vecY_n ) = \vecw_n'  \widehat{\bfSigma}_n^s  \vecw_n.
\]
Estimating $ \widehat{\bfSigma}_n^s $ using the above lower and upper bounds, we obtain the lower bound
\begin{equation}
\label{VarLower}
  \widehat{\Var}( \vecw_n' \vecY_n )_L = \vecw_n'  \widehat{\bfSigma}_n^s  \vecw_n - z_{p} \frac{\widehat{\sigma}_{tr}}{\sqrt{n}}  \| \vecw_n \|_2^2
\end{equation}
and the upper bound
\begin{equation}
\label{VarUpper}
 	\widehat{\Var}( \vecw_n' \vecY_n )_U = \vecw_n'  \widehat{\bfSigma}_n^s  \vecw_n + z_{p} \frac{\widehat{\sigma}_{tr}}{\sqrt{n}}  \| \vecw_n \|_2^2.
\end{equation}
Here $ p = 1 - \alpha/2 $ or $ = 1-\alpha $ if one considers only one of those bounds.

\begin{remark} The normal approximation (\ref{GaussApprox}) holds true under weaker conditions. In particular, the coefficients of the time series may depend on $n$ and are only required to satisfy the weaker decay condition
	\begin{equation}
	\label{DecayWeak}
	\sup_{n \ge 1} \max_{1 \le \nu  \le d_n} | c_j^{(\nu)} | \le C j^{-3/4-\theta/2},
	\end{equation}	
	for some $ \theta \in (0,1/2) $,
	 and the innovations are only required to have finite absolute moments of the order $ 4+\delta $ for some $ \delta > 0 $,
	 see \cite[Theorem 2.3]{StelandSachs2017b}.
\end{remark}

\subsection{Shrinking towards a diagonal matrix}
\label{ref: new}

Let us now study the more general situation to shrink the covariance matrix towards the diagonal matrix.
Here we consider the $d$-dimensional subspace
\[
\calV = \{ \text{diag}( \lambda_1, \dots, \lambda_d ) : \lambda_1, \dots, \lambda_d \in \R \}
\]
which is spanned by the $d$ orthonormal matrices $ \text{diag}( \vece_1 ), \dots, \text{diag}( \vece_d ) \in \calM $, where $ \vece_1, \dots, \vece_d $ are the unit vectors of $ \R^d $ and for a vector $ \veca \in \R^d $. Here and in the sequel we write $ \text{diag}( \veca ) $ for the $ d \times d $ matrix whose diagonal is given by $ \veca $ and all other elements are zero. Further, for a square matrix $ \matA $ we write $ \text{diag}( \matA) $ for the (main) diagonal represented as a column vector and let  \[ \text{diag}^2( \matA ) = \text{diag}( \text{diag}( \matA ) ) = \left( \begin{array}{ccccc} a_{11} & 0 & \cdots & & 0 \\ 0 & a_{22} & 0 & \cdots & 0 \\ \vdots & & & & \vdots \\ 0 & 0 & 0 & \cdots & a_{dd} \end{array}  \right). \] 

The orthogonal projection $ \Pi( \cdot; \calV )  $ onto $ \calV $ is given by
\[
\Pi( \matA; \calV ) = \sum_{j=1}^d (\matA, \text{diag}( \vece_j ) )  \text{diag}( \vece_j ).
\]
Consequently, the optimal shrinkage target is
\[
\bm D = \Pi( \bfSigma_n; \calV ) = \text{diag}( \sigma_1^2, \dots, \sigma_d^2 ).
\]
We estimate $ \matD $ by
\[
\widehat{\matD}_n =  \text{diag}( s_{n1}^2, \dots, s_{nd_n}^2 ),
\]
where $ s_{n1}^2, \dots, s_{nd_n}^2 $ denote the elements on the diagonal of the sample covariance matrix $ \widehat{\bfSigma}_n $.
The corresponding shrinkage covariance estimator is given by
\[
\widehat{\bfSigma}_n^s( \widehat{\matD}_n )  =  (1-W) \widehat{\bfSigma}_n + W \widehat{\matD}_n.
\]

The following new result provides the asymptotics of $ \widehat{\matD}_n $. Recall that
\[
\sigma_\nu^2 = \Var( Y^{(\nu)} ) = ( \bfSigma_n )_{\nu,\nu} 
\]
is the $ \nu$th diagonal element of $ \bfSigma_n $, $ \nu = 1, \dots, d_n $, and observe that
\[
\sqrt{n/d_n} ( \widehat{\matD}_n - \text{diag}^2( \bfSigma_n ) ) = \sqrt{n/d_n} \text{diag}( s_{n1}^2 - \sigma^2_1, \dots, s_{nd_n}^2 - \sigma_{d_n}^2)'.
\]

\begin{theorem} 
	\label{ThNew}
	Assume model (\ref{ModelTS}) with coefficients $ c_j^{(\nu)} $ satisfying the decay condition (\ref{DecayWeak}). Let $ \{ \vecv_n : n \ge 1  \} $ and $ \{ \vecw_n : n \ge 1  \} $ be two sequences of weighting vectors with $ \vecv_n, \vecw_n \in \R^{d_n} $ and \[ \sup_{n \ge 1 } \| \vecv_n \|_{\ell_1} < \infty, \sup_{n \ge 1} \| \vecw_n \|_{\ell_1} < \infty. \]  Then one can redefine the vector time series $ \vecY_{n1}, \dots, \vecY_{nn} $ on a new probability space, together with a  $ d_n $-dimensional Gaussian random vector $ \vecB_n = (B_{n1}, \dots, B_{nd_n} )' $ with covariance structure given by
	\[
	\Cov( B_\nu, B_\mu ) = d_n^{-1} \Cov( s_{n\nu}^2, s_{n\mu}^2 ) + o(1) = d_n^{-1} \beta_n^2( \nu, \mu ) + o(1),
	\]
	such that there exist constants $ C_n $ and $ \lambda $ with
	\[
	\left\| \sqrt{n/d_n} ( s_{n1}^2 - \sigma^2_1, \dots, s_{nd_n}^2 - \sigma_{d_n}^2)' - \vecB_n' \right\|_2 \le  C_n n^{-\lambda}, 
	\]
	as $ n \to \infty $, a.s.. Under the additional assumption $ C_n n^{-\lambda} = o(1) $ we may therefore conclude that
	\[
		\left\| \sqrt{n/d_n} ( s_{n1}^2 - \sigma^2_1, \dots, s_{nd_n}^2 - \sigma_{d_n}^2)' - \vecB_n' \right\|_2 = o(1), 
	\]
	as well as
	\begin{equation}
	\label{ApproxMat}
	\left\| \sqrt{n/d_n} ( \widehat{\matD}_n - \operatorname{diag}^2( \bfSigma_n ) )  - \operatorname{diag}( \vecB_n ) \right\|_F = o(1), 
	\end{equation}
	as $ n \to \infty $, a.s..
\end{theorem}

Observe that (\ref{ApproxMat}) represents an approximation in the space of quadratic matrices of dimension $ d_n \times d_n $.
Theorem~\ref{ThNew} suggests the approximation 
\[
  \sqrt{n/d_n} ( s_{n1}^2 - \sigma^2_1, \dots, s_{nd_n}^2 - \sigma_{d_n}^2)' \sim_{approx} 
  N( \vecnull, \widehat{\matC}_n ) 
\]
where 
\[ 
	\widehat{\matC}_n = \left(  d_n^{-1} \widehat{\beta}_n(\nu, \mu)  \right)_{1 \le \nu \le d_n \atop 1 \le \mu \le d_n}
\]
and the estimators $ \widehat{\beta}_n(\nu, \mu ) $ are defined in (\ref{BetaEst}).

\section{Simulations and application to financial data}  
\label{sec: sims}

\subsection{Simulation study}

We conducted simulations, in order to study the accuracy of the confidence interval 
\[
\left[ \text{tr}^*( \widehat{\bfSigma}_n ) - z_{1-\alpha/2} \frac{\widehat{\sigma}_{tr}}{\sqrt{n}}, \text{tr}^*( \widehat{\bfSigma}_n ) + z_{1-\alpha/2} \frac{\widehat{\sigma}_{tr}}{\sqrt{n}} \right]
\]
for the scaled trace in terms of its coverage probability. Of primary interest is the case that the dimension of the vector time series is of the order of the sample size or even larger. For $ d = 500 $ there are $ 125,250 $ covariances $ \beta_n^2( \nu, \mu ) $ which need to be estimated to calculate the estimator $ \widehat{\sigma}_{tr}^2 $. These computations can, however, be easily parallelized.

Vector time series of dimension $ d $ were simulated following a family of autoregressive processes of order $1$,
\[ 
	Y_{t}^{(\nu)} \sim_{approx} AR(1; \rho_\nu ),
\] 
where $ \rho_\nu = 0.1 + (\nu/d) 0.5 $, $ \epsilon_i \stackrel{i.i.d.}{\sim} N(0,1)$, $ 1 \le \nu \le d $. The weights were chosen as
\[
  w_{mh} = \left\{ \begin{array}{cc} 1, \qquad & |h|  \le m+1, \\
  0, \qquad & \text{else},  \end{array} \right.
\]
with lag truncation constant  $ m = \trunc{n^{0.3}} $. The nominal coverage probability was chosen as $1-\alpha = 0.9$. The true value of the scaled trace norm of the corresponding true covariance matrix was estimated by a simulation using $ 20,000 $ runs. Then the coverage probability was estimated by $ 1,000 $ Monte Carlo simulations. The simulations were carried out using R and the doParallel and foreach packages for parallel computations.  

Table~\ref{Tab1} provides the simulated coverage probabilities for sample sizes $ 10, 100, 250 $ and dimensions  $ 10, 50, 100, 250, 500 $. As our theoretical results do not require a constraint on the dimension such as convergence of $ d/n $ to a constant between $0$ and $1$, we simulate all resulting combinations. 
The nominal coverage probability is $ 1-\alpha = 0.95 $. It can be seen that the coverage is good, if the sample size is not too small; especially, the coverage gets better if $n$ increases. It is remarkable that, according to the simulation results, the accuracy is quite uniform in the dimension, even when the dimension is much larger than the sample size as in the case $ d = 500 $ and $ n = 10 $.

\vskip 0.4cm
	\begin{table}
\begin{center}
	\begin{tabular}{c|ccccc} \hline
		$n \backslash d $ & $ 10 $ & $ 50 $ & $ 100 $ & $ 250 $ & $ 500 $ \\ \hline
		$ 10 $ & 0.885 & 0.897 & 0.884 & 0.900 & 0.902 \\
		$ 100 $ & 0.917 & 0.915 & 0.916 & 0.921 & 0.917 \\
		$ 250 $ & 0.932 & 0.935 & 0.930 & 0.913  & 0.932 \\ \hline
	\end{tabular}
\end{center}
\caption{Simulated coverage probabilities of the proposed confidence interval for the scaled trace.}
\label{Tab1}
	\end{table}

\subsection{Application to asset returns}

We applied the proposed methods to three data sets, in order to illustrate their potential benefit in practice. The first one, NYSE, is a standard data set of asset returns from the New York stock exchange used by \cite{Cover1991}, \cite{GLU2006} and others. The NYSE data set includes daily closing prices of $  32 $ stocks over a 22-year period from July 3rd, 1962 to December 31th, 1984. The second one, TSE, consists of returns of  $ 88 $ stocks of the Toronto stock exchange  for the 5-year period from January 4th, 1994 to December 31st, 1998.  The last data set
%, downloaded from https://github.com/CNuge/kaggle-code, 
consists of 470 stocks of the SP500 over the 5-year period from February 8th, 2013 to February 7th, 2018.

In a first experiment, we estimated nonparametrically the $ d \times d $ dimensional covariance matrix of the associated log returns for the first $ 250 $ log returns of the NYSE data set. \cite{LW2004} proposed an estimator of the (optimal) shrinkage weight $ W $ leading to the estimate $ 0.172 $; for the other data sets the estimates are larger.
Hence, in all analyses we use the weight $ W = 0.2 $. In this way we keep the regularization at a moderate level and can mask out effects due to the estimation error with respect to $W$. Further, this also allows better comparisions across the data sets and subsamples.

How does the condition number, defined as the ratio of the largest eigenvalue and the smallest one, improves by shrinking? The following figures provide some insights. When using only the last $ n = 50 $ log returns of the NYSE data set, the condition number of the sample covariance matrix is $ 291.79 $. Shrinking substantially improves the condition number by more than a factor of $10$ to $ 27.84 $. For the TSE stock data with $ d = 88 $ stocks, the condition number decreases from $ 217.7 $ to $ 71.29 $. Lastly, for the SP500 data from 2013-2018 the condition number for the $ 470 \times 470 $-dimensional covariance matrix decreases from  $6,245.4$ to $964.4$.

\begin{figure}
\begin{center}
\includegraphics[width=10cm]{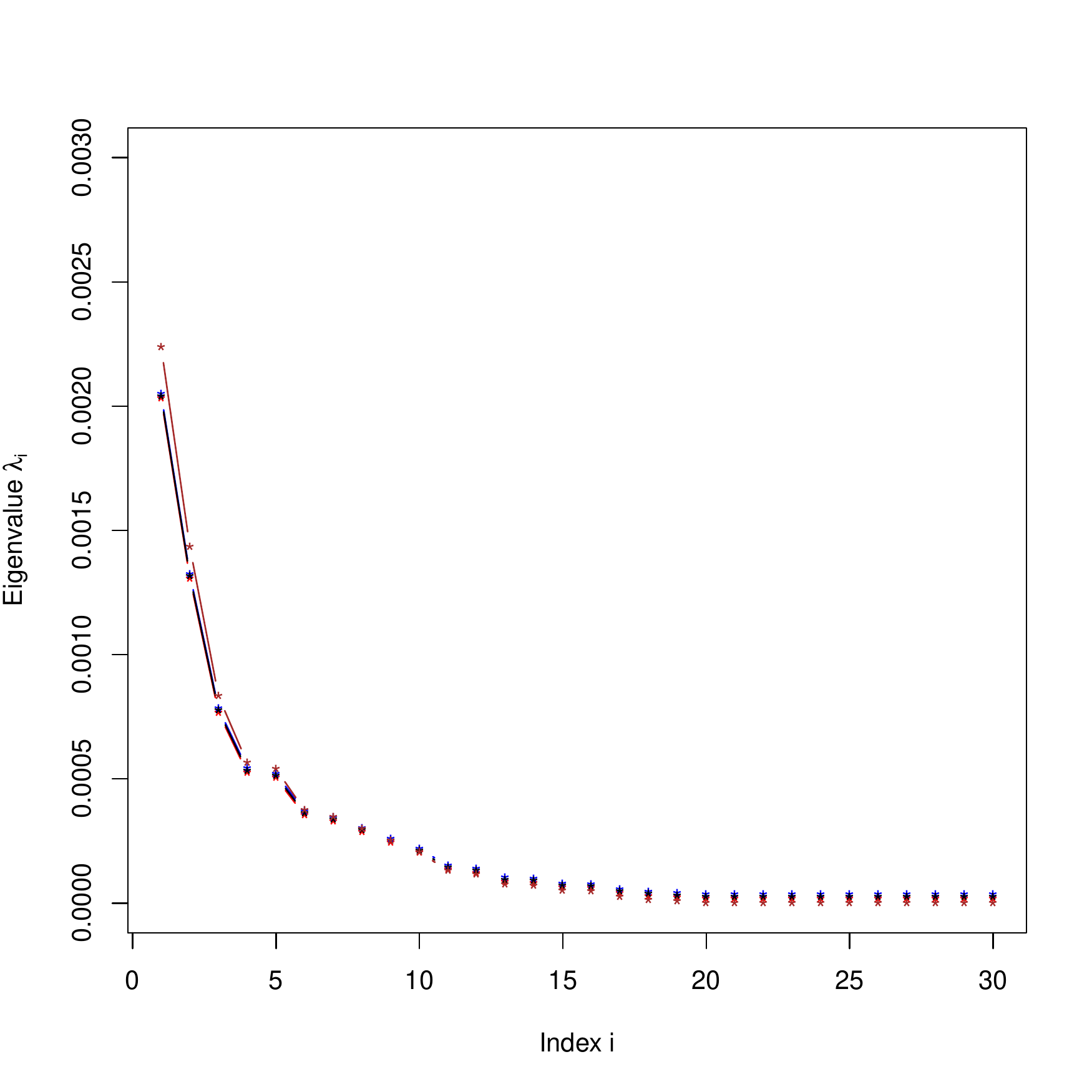}	
\end{center}
\caption{Estimated eigenvalues based on $ n = 50 $ asset returns from $32 $ NYSE stocks: Sample covariance matrix (points), shrinkage estimator as well as eigenvalues of the lower and upper bounds (drawn as error bars) for the shrinkage covariance matrix estimator.}
\label{Fig1}
\end{figure}

For a confidence level of $ 99\% $ the lower and upper bounds for the shrinkage covariance matrix were calculated. We report the eigenvalues as an informative summary statistic. Figure~\ref{Fig1} shows the eigenvalues of the sample covariance matrix $ \widehat{\bfSigma}_{50} $, the shrinkage estimator $ \widehat{\bfSigma}_{50}^s $ and of the lower and upper bounds for the NYSE data. One can see that the eigenvalues of the lower (upper) bound are always smaller (larger)  than eigenvalues of the shrinkage estimator. However, note that the corresponding intervals can not be interpreted as confidence intervals. 

\begin{figure}
	\begin{center}
		\includegraphics[width=10cm]{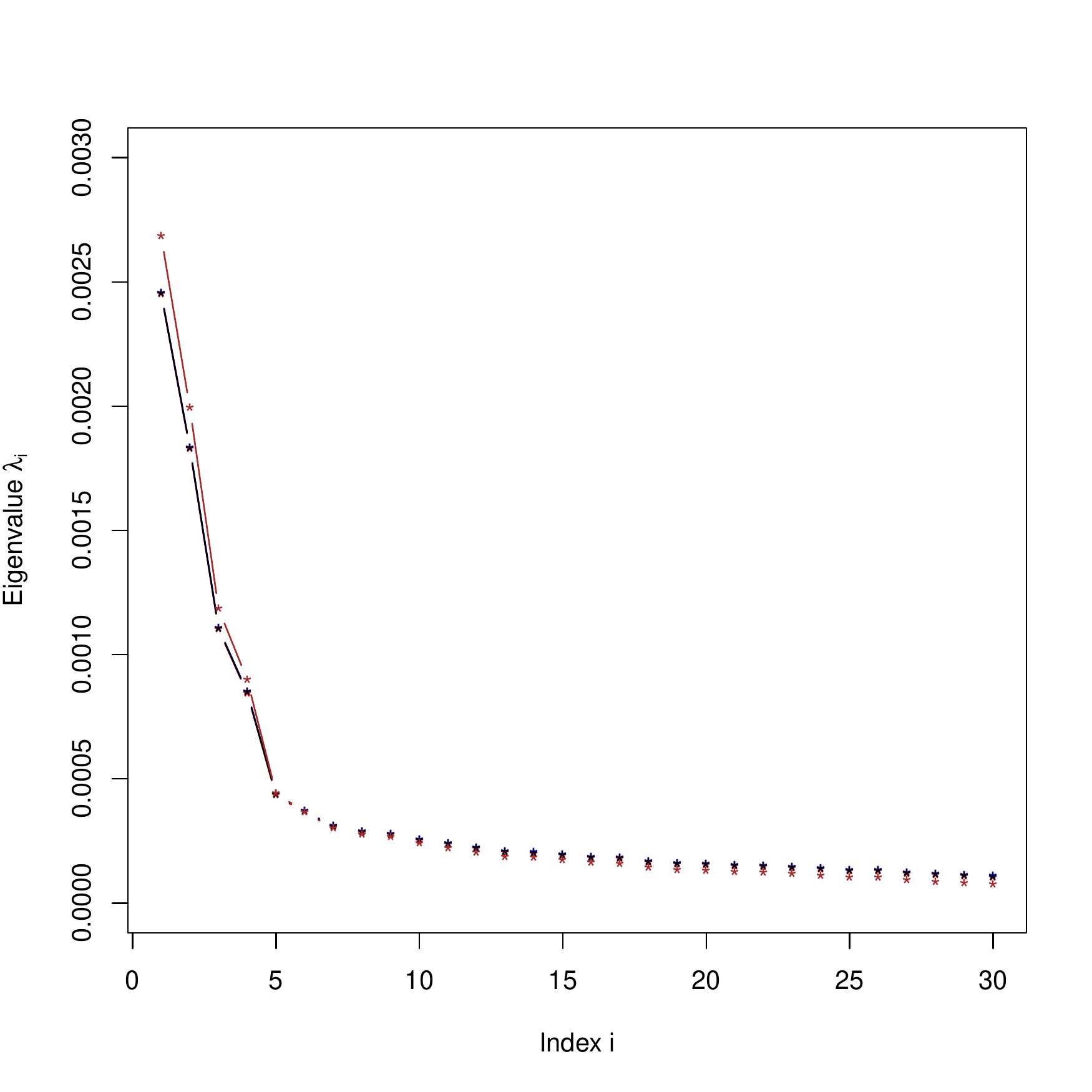}
	\end{center}
\caption{Estimated eigenvalues based on $ n = 5,651 $ asset returns from $32$ stocks: Sample covariance matrix (brown) and shrinkage estimator (black).}
\label{Fig2}
\end{figure}

As a comparison, Figure~\ref{Fig2} provides the eigenvalues of the sample covariance matrix and the shrinkage estimator when using the full data set of $ n = 5,651 $ trading days.

In addition to the above analysis, we applied the proposed lower and upper bounds to portfolio optimization. Recall that the classical approach to porfolio optimization is to minimize the portfolio variance under the constraint $ \vecw_n' {\bm 1} = 1 $ and, optionally, constrained on a specified mean portfolio return. Here and in what follows, we focus on the variance-minimizing portfolio, called minimum variance portfolio, $ \vecw_n^* $ which minimized the variance under the constraint $ \vecw_n' {\bm 1} = 1 $,
\[
  \vecw_n^* = \argmin_{\vecw_n: \vecw_n' {\bm 1} = 1} \vecw_n' \bfSigma_n \vecw_n, 
\]
where $ \bfSigma_n $ is the true covariance matrix of $n$ daily log returns. If there are no short sales in the optimal portfolio, its $ \ell_1 $-norm is $1$. For real markets, this condition usually does not hold true. Then we need to assume that $ \sup_{n \ge 1} \| \vecw_n^* \|_{\ell_1} < \infty $. This can be guaranteed by adding an appropriate penalty term to the optimization problem as, e.g., in \cite{BD2009}, which often leads to $ \ell_0 $-sparse portfolios, i.e. one holds only positions in a subset of the available stocks. Nevertheless, we stay with the variance-minimizing portfolio in our analysis, which holds (long or short) positions in all assets, so that all covariances between the asset log returns are relevant to calculate the variance estimator $ \wh{\sigma}_{tr}^2 $. For $\ell_0 $-constrained portfolios holding positions only in a subset of the assets one can presumably expect tighter bounds for the portfolio risk than reported here for the variance-minimizing portfolio.

%When using the estimator proposed by \cite{Ledoit2003}, the weights could 
The whole data set was split in subsamples of $ n = 252 $ returns corresponding to one year. For each year $t$ the optimal portfolio $ \vecw_{nt}^* $, its associated estimated risk $ \sqrt{ \vecw_{nt}^*{}' \wh{\bfSigma}_n^{s} \vecw_{nt}^* } $ and the lower and upper bounds 
\[
 \sqrt{ \vecw_n^*{}'  \widehat{\bfSigma}_n^s  \vecw_n^* \pm z_{p} \frac{\widehat{\sigma}_{tr}}{\sqrt{n}}  \| \vecv_n^* \|_2^2 }
\]
for $ p = 1-0.995 $ were calculated, cf. (\ref{VarLower}) and (\ref{VarUpper}). If the expression under the square root is negative, then it is set to $0$.

\begin{figure}
	\begin{center}
		\includegraphics[width=10cm]{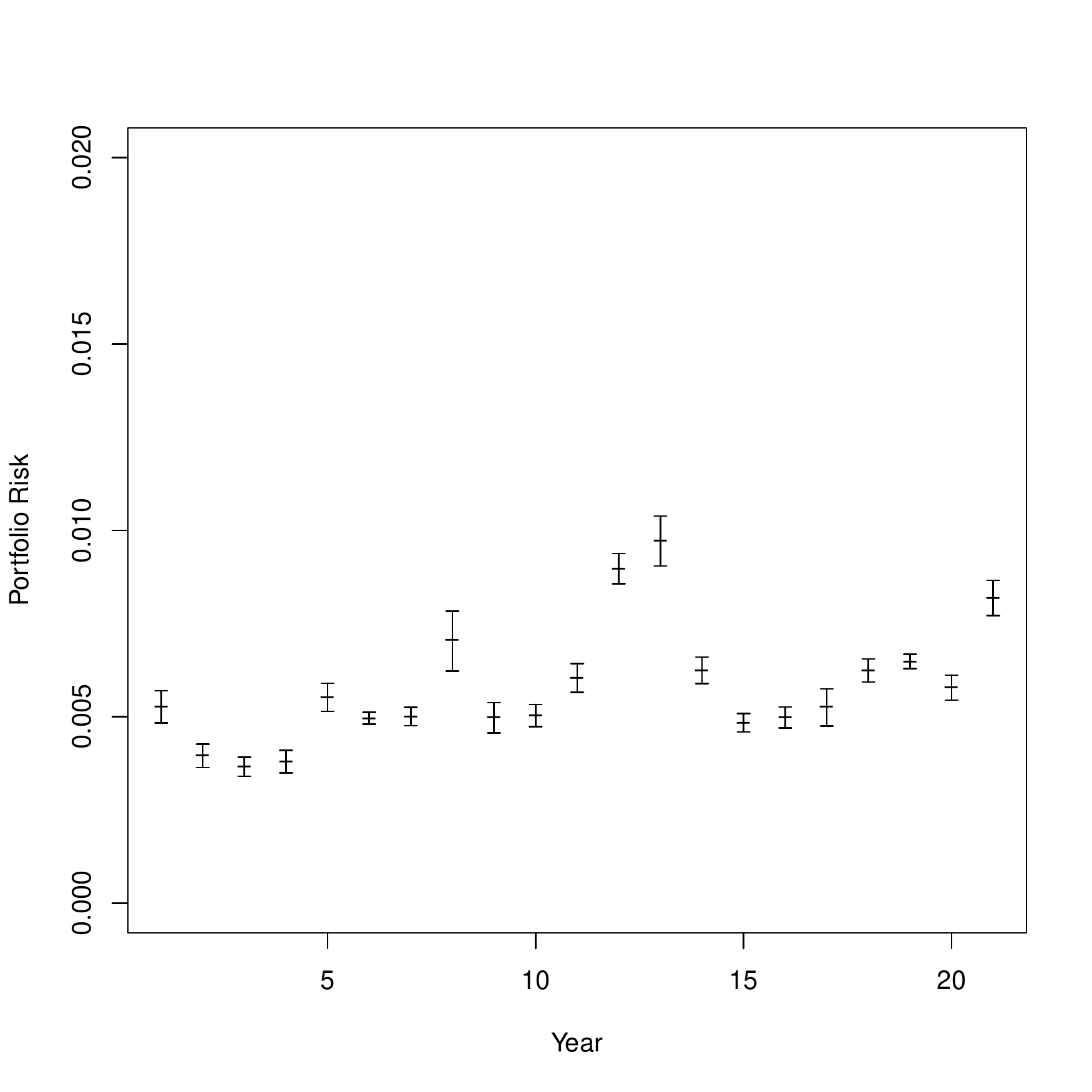}	
	\end{center}
	\caption{Yearly estimated portfolio risk as well as lower and upper bounds of $32$ stocks of the New York stock exchange over $22$ years from 1962 to 1984. The optimal portfolio is calculated using the shrinkage covariance estimator.}
	\label{Fig3}
\end{figure}

\begin{figure}
	\begin{center}
		\includegraphics[width=10cm]{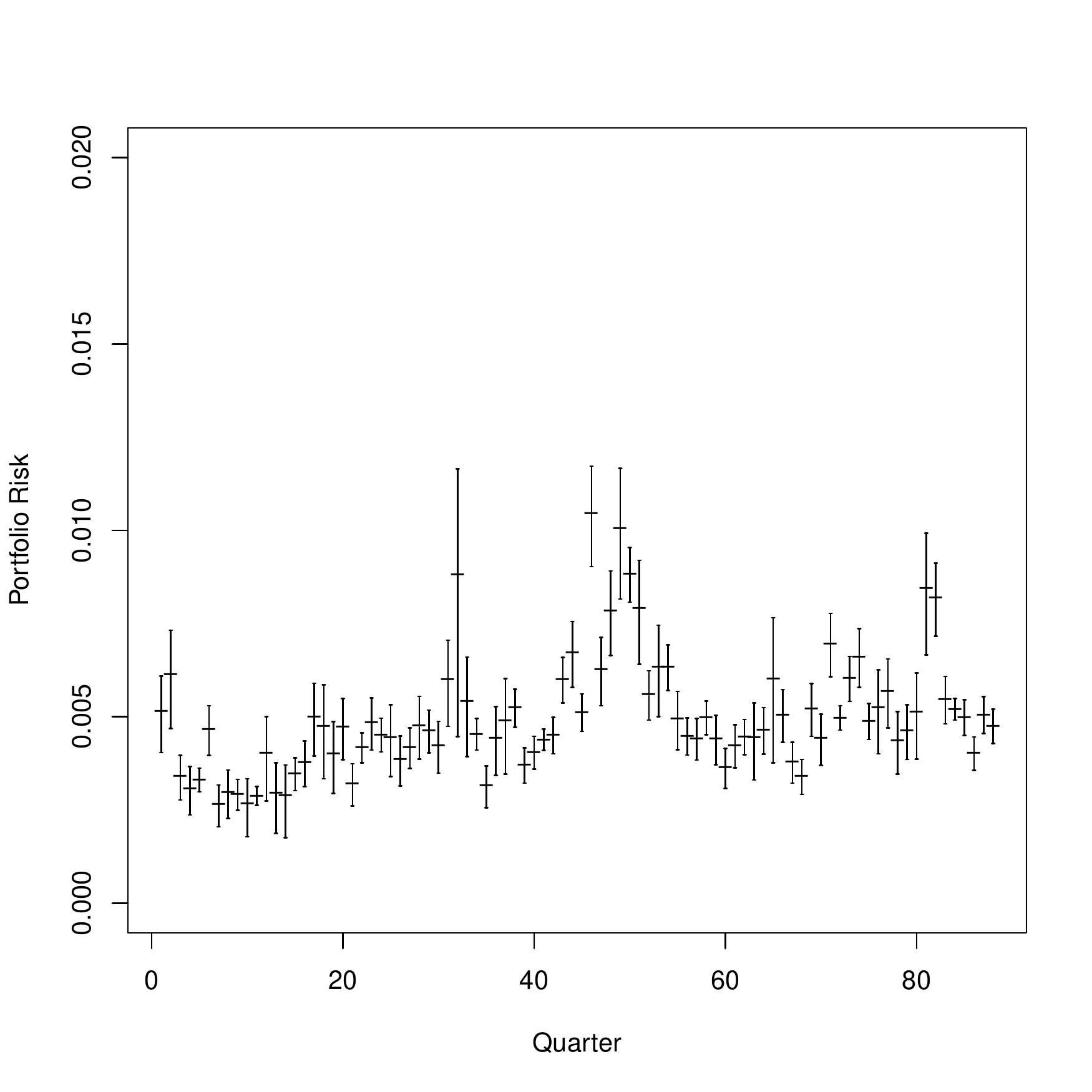}	
	\end{center}
	\caption{Quarterly estimated portfolio risk as well as lower and upper bounds of $32$ stocks of the New York stock exchange over $22$ years (= $88$ quarters) from 1962 to 1984. The optimal portfolio is calculated as in Figure~\ref{Fig3}.}
	\label{Fig4}
\end{figure}

Figure~\ref{Fig3} shows the result for the NYSE data set. This analysis was repeated on a quarterly basis based on $ n = 63 $ trading days. The result is shown in Figure~\ref{Fig4}. One can observe that the bounds are less tight due to the smaller sample size available for estimation which increases the statistical estimation error.

\begin{figure}
	\begin{center}
		\includegraphics[width=10cm]{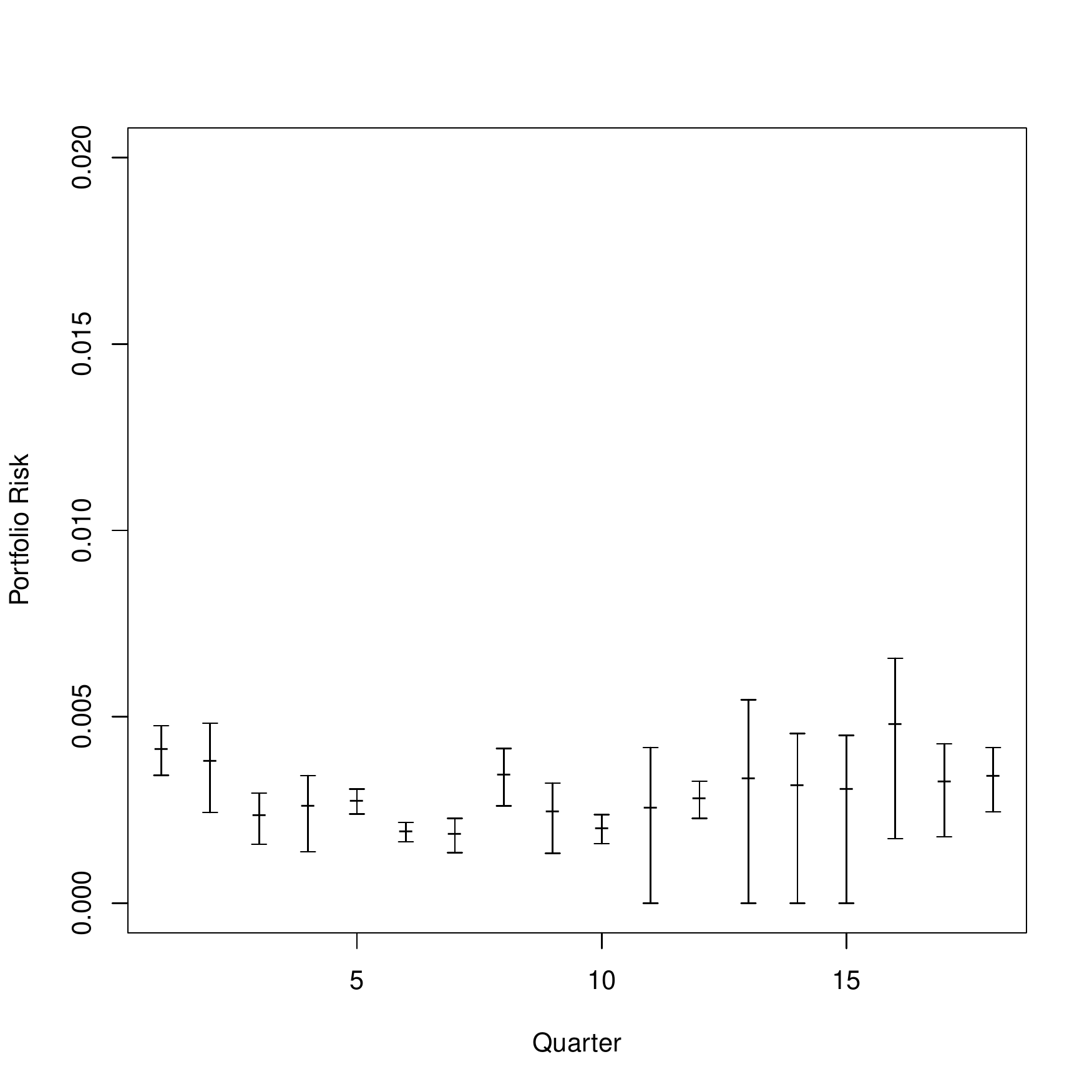}	
	\end{center}
	\caption{Quarterly estimated portfolio risk as well as lower and upper bounds for $88 $ stocks of the Toronto stock exchange over $5$ years. The optimal portfolio is calculated using the shrinkage covariance estimator.}
	\label{Fig5}
\end{figure}

For the TSE data set where $ d = 88 $ the corresponding results on a quarterly basis are shown in Figure~\ref{Fig5}. Here the covariance matrix is estimated based on $ n = 63 $ log returns, such that $ d > n $. 

Lastly, for the SP500 data set for the period from February 2013 to February 2018,  Figure~\ref{Fig5} shows the corresponding portfolio risks and their bounds on a quarterly basis. Here $ 63 $ return vectors are used to estimate the $ 470 \times 470 $-dimensional covariance matrix and the associated risks of the variance-minimizing portfolio. 

\begin{figure}
	\begin{center}
		\includegraphics[width=10cm]{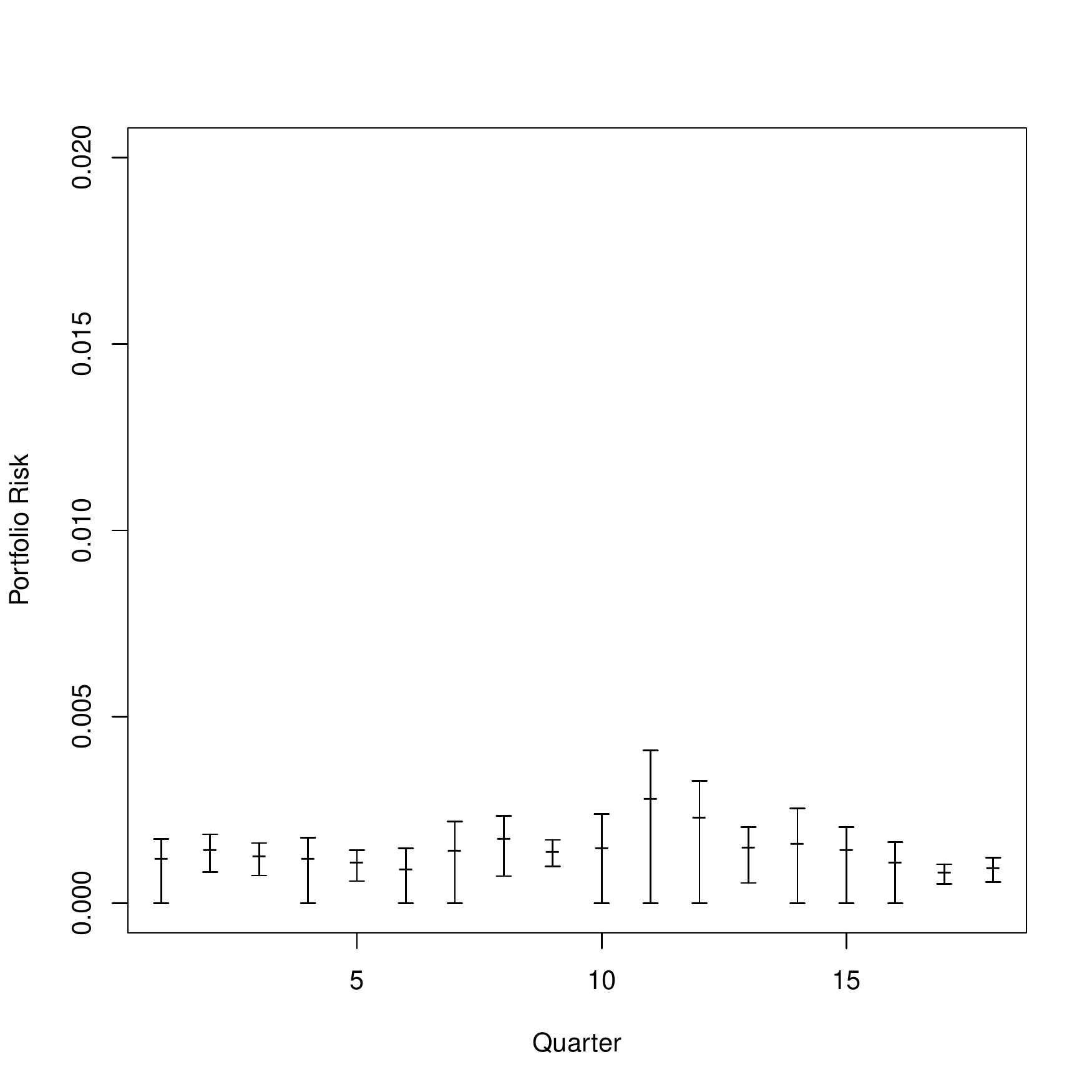}	
	\end{center}
	\caption{Quarterly estimated portfolio risk as well as lower and upper bounds for $470 $ stocks of the SP500 index over the $5$-year-period from 2013 to 2018. The optimal portfolio is calculated using the shrinkage covariance estimator.}
	\label{Fig6}
\end{figure}

%\subsubsection{Confidence intervals for portfolio risks}

%Let us denote the vector of daily log returns by $ {\bm R}_t $. For a portfolio vector $ {\bm w}_t $ the associated portfolio risk can be defined as $ \Var( {\bm w}_t' {\bm R}_t ) = {\bm w}_t' {\bm \Sigma}_t {\bm w}_t $ or $ \sigma_t( {\bm w}_t ) = \sqrt{{\bm w}_t' {\bm \Sigma}_t {\bm w}_t } $. Assuming the required assumptions are satisfied, the proposed confidence interval for $ \Var( {\bm w}_t' {\bm R}_t ) $ provides us with a confidence interval for $ \sigma_t( {\bm w}_t ) $, namely \[ \left[ \sqrt{\text{tr}^*( \widehat{\bfSigma}_n ) - z_{1-\alpha/2} \frac{\widehat{\sigma}_{tr}}{\sqrt{n}}}, \sqrt{\text{tr}^*( \widehat{\bfSigma}_n ) + z_{1-\alpha/2} \frac{\widehat{\sigma}_{tr}}{\sqrt{n}}} \right]. \] Here...

%      [,1]  [,2]  [,3]  [,4]  [,5]  [,6]
%[1,] 0.859 0.850 0.868 0.860 0.859 0.847
%[2,] 0.862 0.885 0.873 0.874 0.885 0.873
%[3,] 0.885 0.878 0.882 0.863 0.880    NA

\section*{Acknowledgments}

This work was supported by a grant from Deutsche Forschungsgemeinschaft, grant STE 1034/11-1. Comments from anonymous reviewers are appreciated.

%\bibliographystyle{plain}
%\bibliography{lit}

\appendix 

\section{Proof of Theorem~\ref{ThNew}}

We give a sketch of the proof. We apply \cite[Theorem~2.3]{StelandSachs2017b}, which generalizes \cite{StelandSachs2017a} and is based on techniques of \cite{Kouritzin1995} and \cite{Ph1986}. \cite[Theorem~2.3]{StelandSachs2017b} is applied with $ \vecv_n^{(j)} = \vecw_n^{(j)} = \vece_j $, where $ \vece_j $ denotes the $j$th unit vector of the Euclidean space $ \mathbb{R}^{d_n} $, $ j = 1, \dots, d_n $ and $ L_n = d_n $. Basically, the result asserts that, on a new probability space, one can approximate the partial sums \[ D_{nk}^{(j)} = \sum_{i=1}^k \vecv_n^{(j)}{}' \vecY_{ni}  \vecw_n^{(j)}{}' \vecY_{ni}, \qquad j = 1, \dots, L_n, 1 \le k \le n, n \ge 1 \] 
by a Brownian motion, and here the number of such bilinear forms $ L_n $ ($= d_n$ in our case) given by $L_n $ pairs of weighting vectors, may grow to $ \infty $. Using the notation and definitions of Secton 2.3 of \cite{StelandSachs2017b}, we obtain
	\begin{align*}
	\calD_{nj}(1) &= \frac{1}{\sqrt{n d_n}} D_{nn}^{(j)} ( \vecv_n^{(j)}, \vecw_n^{(j)}  )  \\
	& =  \frac{1}{\sqrt{n d_n}} \vece_j'( n \widehat{\bfSigma}_n - \E(n \widehat{\bfSigma}_n) ) \vece_j \\
	& = \sqrt{n/d_n} \left( s_{nj}^2 - \sigma^2_j \right)_{j=1}^{d_n}.
	\end{align*}
	By \cite[Theorem~2.3]{StelandSachs2017b} we may redefine the above processes, on a new probability space together with a Gaussian random vector $ \vecB_n = (B_{n1}, \dots, B_{nd_n} )' $ with $ \E( \vecB_n ) = 0 $ and covariances given by
	\[
	\Cov( B_{n\nu}, B_{n\mu} ) = d_n^{-1} \widetilde{\beta}_n^2( \nu, \mu ),	  
	\]
	where $ \widetilde{\beta}_n^2( \nu, \mu ) $ satisfy
	\[
	d_n^{-1} |\widetilde{\beta}_n^2( \nu, \mu ) - \beta_n^2( \nu, \mu ) | = o(1),
	\] 
	by Lemma~2.1 and Theorem~2.2 of \cite{StelandSachs2017b} (there the quantities $ \widetilde{\beta}_n^2( \nu, \mu ) $ are denoted by $ \beta_n^2( \nu, \mu )$), such that the strong approximation 
	\[
	\sum_{j=1}^{d_n} \left| \calD_{nj} - B_j  \right|^2 = o(1),
	\]
	as $n \to \infty $, a.s., holds true. But this immediately yields
	\[
	  \left\| \bigl( \sqrt{n/d_n} ( s_{nj}^2 - \sigma^2_j ) \bigr)_{j=1}^{d_n} - \vecB_n \right\|_2 = o(1),
	\]
	as $ n \to \infty $, a.s.. Now both assertions follow, because
	\[
	\left\| \sqrt{n/d_n} ( \widehat{\matD}_n - \text{diag}^2( \bfSigma_n ) )  - \text{diag}( \vecB_n ) \right\|_2 = \left\| \bigl( \sqrt{n/d_n} ( s_{nj}^2 - \sigma^2_j ) \bigr)_{j=1}^{d_n} - \vecB_n \right\|_2.
	\]	
	$ \hfill \Box $

\end{document}